\documentclass{article}
\newcommand{\be}{\begin{equation}}
\newcommand{\ee}{\end{equation}}
\begin{document}

\baselineskip20pt

\title{Signatures of Randomness in Quantum Chaos}

\author{Piotr Garbaczewski\thanks{Presented at the XIV Marian Smoluchowski Symposium
on Statistical Physics, Zakopane, Poland,  September 9-14, 2001}\\
Institute of  Physics, University  of Zielona G\'{o}ra, pl. S{\l}owia\'{n}ski  6\\
PL-65 069 Zielona G\'{o}ra,  Poland \\}
\maketitle

\begin{abstract}
We investigate toy dynamical models  of energy-level repulsion in  quantum
eigenvalue  sequences. We focus on parametric (with respect to a running
coupling or "complexity" parameter) stochastic processes that are
capable of relaxing towards a stationary regime  (e. g. equilibrium, invariant
asymptotic measure).  In view of  ergodic property, that  makes
them appropriate for  the study of short-range fluctuations in any  disordered,
randomly-looking spectral sequence (as exemplified  e. g. by   empirical
nearest-neighbor spacings histograms of various  quantum systems).
The pertinent Markov diffusion-type
processes (with values in the space of  spacings) share a  general form of
forward drifts $b(x) = {{N-1}\over {2x}} - x$, where $x>0$ stands for the spacing value.
Here $N = 2,3,5$ correspond  to the familiar (generic)
random-matrix theory inspired cases, based  on the exploitation of
the Wigner surmise (usually regarded  as an approximate  formula).
 $N=4$ corresponds to the (non-generic) non-Hermitian Ginibre
 ensemble. The result appears to be  exact in the context of $2\times 2$
 random matrices and indicates a potential validity of  other non-generic
 $N>5$ level repulsion laws.
\end{abstract}

\vskip1.0cm

\section{Regular versus irregular in quantum chaos}

The vague notion of so-called quantum chaos,  normally  arising  in
conjunction with    semiclassical  quantum mechanics of
chaotic  dynamical systems  \cite{stock}, currently  stands for
 a key-word capturing   continued efforts  to give
a  proper account to what  extent  quantization destroys,
preserves,  or qualitatively reproduces  major  features of
classical chaos.
There is no  general agreement about
  what actually \it  is \rm   to be interpreted as  "quantum chaos" or
  its definite
 manifestations.
That in part derives from   an inherent ambiguity  of quantization schemes
for nonlinear, possibly nonconservative,  driven and damped classical
problems,  and is intrinsically   entangled with a  delicate reverse
problem of a reliable (semi)classical limit procedure for
 once given quantum system.
 Other   origins  of this elusiveness seem to be
 rooted  in the   diversity of meanings attributed  in the literature
to the concepts  of    \it regular \rm and/or \it irregular \rm  behaviour
of a physical system, irrespective  of whether it is classical or quantum.

Mathematical definitions  of classical chaos emphasize   an apparent
appearance of \it randomness   \rm in deterministic situations,
\cite{lieberman,karp}.  That  involves  a deep  question  of when specific
features of a physical system appear to be (or can be interpreted as) random.
Quite typically, while dealing  with  an irregular
behaviour, we need to quantify an interplay  between chance and order in
terms of suitable  measures of   randomness (if random, then "how much"?),
\cite{beltrami,knuth,fishman}.

Disorder, irregularity  and randomness are  casually perceived as synonyms
 and are interpreted to stay  at variance with   notions of order and
 regularity. Albeit order and randomness may as well  coexist as
 "two faces of the same mysterious coin", \cite{beltrami}.

One of basic  problems in the quantum chaos theory is to establish whether
 the classical  order - disorder interplay   induces any
 unambiguous imprints ("signatures of chaos", \cite{haake}) in
  quantum systems.
In this context,  familiar  concepts   of  regular and irregular spectra
 \cite{percival,berry,pechukas} were coined  to characterize
 distinctive differences between   semiclassical distributions of energy
 eigenvalues  for \it generic \rm quantum systems.
 The term "generic" basically  means "more or less typical", since
 one  excludes from considerations all systems which do not behave \it properly, \rm
 although  there are  many of them. (More stringent definition invokes suitable
 symmetry properties of the quantum system.)

 Nowadays it is clear that  an  irregularity alone
 of  any  particular spectral series (possibly  interpreted  in terms of an
 irregular sequence of consecutive  enery or quasi-energy levels)   is not
 an adequate criterion for quantum manifestations of chaos.  In fact,
semiclassical  spectra  corresponding to  many  \cite{berry}
classical systems, be them
  integrable or chaotic,  have an irregular appearance.  That
 was the  motivation for attempts to   classify such spectra in terms of  the
  "degree of randomness of the sequence of eigenvalues", \cite{guarneri}, cf.
 also  \cite{knuth} and \cite{fishman} for related argumentation.

  Let us however stress  that  a primordial question of whether  a given energy
  level sequence can at all be regarded as random  has been left untouched.
  Seldom one  may have at disposal  a complete  analytic information about
   quantum spectra. Usually some experimentation is needed to
   extract the data and
    most of available  spectral information  relevant for quantum chaos studies,
    comes either from a genuine  experiments (microwave analogs of
    quantum billiards, realistic nuclear
  data) or computer simulations,  always   with a definite  beginning and  an end.
  In each case that  produces  a finite string  of data and it is known that no finite
  sequence can  be  interpreted as truly random.
    Fortunately,  if  a  data sequence   generated
  by  a stochastic process of any origin (deterministic algorithms included)
 is sufficiently long, then it will always satisfy a test for randomness
 with  fine-tuned confidence level, cf. \cite{knuth,lof,silverman}.

We emphasize an issue of randomness,  because various probability laws (and
densities of invariant measures) are omnipresent in the quantum chaos
research. In view of that,  a stochastic modeling will be our major tool in below.

  A rich class of  classically  integrable  (hence regarded as  regular)
  systems displays random -looking, locally uncorrelated sequences of energy
  eigenvalues,  \cite{berry,farrelly,berry1}. However there  are well known
  classically chaotic (hence  regarded as irregular) systems whose quantum spectral
  statistics appears to bear no distinctive imprints of classical chaos
  and  look  appropriate for the  completely  integrable case,
   \cite{farrelly}.
 Therefore, a supplementary rule is necessary to typify various
 classes of spectral irregularities and of the involved  types of randomness
(being random, but possibly   "random otherwise"), if those are to be interpreted as
consequences of   irregular characteristics of the related classical system.

 A  possible hint might have  originated from  discriminating
 between the spatial regularity and
 irregularity of the corresponding quantum  eigenfunctions. It is
 the spatial pattern of wave functions that appears to  have  a decisive effect
 on   the spectral pattern of eigenvalues,
 \cite{pechukas,berry1}.
However, a minor obstacle still  persists: not all classically ergodic
systems (irregular case of Ref. \cite{percival}) would semiclassically
yield  irregular eigenfunctions, \cite{pechukas}.
Consequently, one usually  tries to narrow the class of quantum system  that
are suspected to show    undoubtful  "signatures of chaos" to those which
have irregular eigenfunctions, with  no specific reference to their
classical (chaotic or non-chaotic) behaviour. In this  class one   ultimately
attempts to  specify those systems which remain  in a  consistent  semiclassical
relationship with their  chaotic  classical partners. Those
systems quite justifiably  would deserve to be named  generic and
would  more or less naturally fall into various spectral
universality classes, in accordance with the random-matrix
classification scheme, \cite{mehta,brody,bohigas}.

Under rather plausible assumptions, \cite{pechukas},  quantum
systems with  spatially irregular wave functions were found to
exhibit \it level repulsion\rm, hence to "avoid" degeneracies
which is basically an indication of non-integrability, hence  not
 necessarily  that of any links with chaos.

We  recall that an  opposite spectral effect of level clustering,
combined with the conjectured absence of correlations between
levels, is characteristic for a large class ("almost all"
according to \cite{berry1,veble}, see however \cite{marklof}) of
classically integrable systems.
Typically they  display so-called  Poisson statistics (strictly
speaking, there is an exponential law of probability involved
\cite{feller}) of adjacent level spacings: small spacings are
predominant and there is enough room for multiply degenerate
levels.  One says then \cite{rosenzweig} that energy levels occur
in a completely random way via a Poisson process on the energy axis.

Therefore,  \it level repulsion, \rm  when regarded as an  \it
emergent \rm spectral symptom of level correlations (usually interpreted as a
 certain  departure from purely random behaviour),  may be viewed as a
 necessary condition to
deal either with  quantum  imprints of classical chaos or, in the  least,
with  a classically nonintegrable phase-space irregularity.

Interestingly enough, this viewpoint finds some support in
the discovery of  pseudointegrable systems (variously shaped billiards,
sometimes with singular scattering obstacles) which appear to be neither
integrable nor chaotic, but give rise to various forms of "wave chaos"
while quantized, \cite{richens,seba}.
The corresponding distribution  of adjacent level spacings is named  semi-Poisson
and combines various forms (including fractional powers) of  level repulsion with
Poisson (exponential) statistics, hence purely random behaviour in the spectral series.
The  repulsion  phenomenon is here a
manifestation of the topologically complicated phase space (an invariant manifold
 is  not topologically equivalent to a torus but to a higher genus manifold), which
was conjectured to preclude integrability and thus the standard torus (EBK)
semiclassical quantization,  see e.g. \cite{richens}-\cite{bogomolny}.

In view of the  wide usage of such terms like "universality"   in
the quantum chaos literature, one  should always keep in mind that  harmonic
oscillators display level repulsion, \cite{berry},
although they seem to be exemplary cases of classical  and quantum regularity
at their   extreme. Another spectacular exception is the hydrogen atom spectrum.
Like all higher dimensional harmonic oscillators, or a square billiard
\cite{keating,veble},  the Coulomb spectrum   belongs to a distinctive group
of "pathologically nongeneric" spectral problems, \cite{haake}.

Nonetheless we shall confine  our attention to   the  suggestive,
random matrix theory universality  classification  that is considered
to be faithful for local fluctuations in  quantum
spectra of (generic) systems that display global chaos in their classical
phase spaces.  Our hunch is to mimic (or rather extract) those features of the
level-spacing classification which  may bear imprints of pure randomness or
in reverse - depart from randomness.

Studying classical manifestations of chaos in terms of probability measures
(including their densities or distributions and their dynamics) is
a respectable strategy, \cite{lasota}.
In quantum theory, in  view of Born's  statistical interpretation postulate alone,
probability measures are  ubiquitous.
On the other hand, various probability laws and distributions
 naturally pervade the  familiar
random-matrix theory  \cite{mehta,brody}. This  statistical theory of spectra,
 models  a symmetry -limited spectral disorder in terms of
statistical ensembles of complex quantum systems (e.g. heavy nuclei). Apart from
  an  ensemble input, random-matrix theory  forms  a convenient vehicle to  interpret
spectroscopic properties of a concrete (single !)  quantized version  of  a
 complex  classical model.
(The classical complexity  notion
  refers e.g. to the  phase-space  organization specific to a system  and  various
  complication degrees of its dynamics related to
   ergodicity, mixing and  exactness.)

However, one should keep  in mind that the universality  hypothesis
in the context of quantum chaos proper, derives from  exploiting  a   spectral
affinity   of an  \it ensemble \rm of large (with size ultimately  growing
to infinity) random matrices,  with a once given \it  individual  \rm quantum  system
(take into consideration the Sinai billiard or (a)periodically kicked pendulum/rotator).
Therefore, we may justifiably ask how an individual (Hamiltonian or Floquet-type)
quantum eigenvalue problem  may capture   all conceivable statistical properties of suitable
random-matrix   ensemble spectra ?  Told otherwise, how may we justify a  comparison of
a statistical ensemble of  disordered spectral series   with   \it  one \rm
only  specific  energy (or quasi-energy) level sequence  of an a
priori chosen quantum system  ?

To our knowledge this  conceptual obstacle, except for preliminary
investigations of Ref. \cite{valz},  has not  received much attention
in the quantum chaos literature.
A  partial answer to that  question, \cite{haake},
points towards certain \it
ergodicity \rm properties appropriate for   models of the  parametric
level dynamics (Coulomb gas, plasma or else, evolving in "fictitious time"),
 that  provide
a reinterpretation of random-matrix theory in terms of an equilibrium
statistical mechanics for a  fictitious $N$-particle  system
(with $N$ allowed to grow indefinitely).

In the framework of random-matrix theory, an  ergodic problem for Gaussian
 ensembles was analyzed  long ago in Ref. \cite{pandey},
with a focus on the ergodic behaviour  for the
eigenvalue density and  $k$-point correlation functions of
individual random matrices and their statistical  ensembles. That  involves
a local version of the ergodic theorem,  where e.g. the spectral averaging over
a finite energy span of the level density is compared with the matrix
ensemble mean of the  level density.
 That suggests analogies with disordered  many-body quantum systems
where ensemble averaging is a standard analytic tool, while for
an individual system, only an energy averaging should be employed,
\cite{altshuler}.

We know that the distribution of spacings of highly excited quantum systems may involve
definite laws  like e.g. the exponential or Wigner-type distributions.
Such laws may be related to definite stochastic processes as invariant measures,
 in particular as asymptotic measures to which the process does relax.

 We shall focus upon the  parametric dynamics (parametric interpolation)
scenario for  the nearest-neighbor spacing distributions  of   irregular quantum
systems where asymptotic invariant (with respect to the parametric process)
probability measures are ultimately involved.
(Let us mention that a concept of parametric dynamics involves the possibly troublesome
"fictitious time" parameter. Its possible interpretation is that of  a running
coupling constant measuring the strength of the chaotizing perturbation,
 or more generally that of a "complexity parameter" whose growth to infinity gives
 account of the complexity increase in the spectral properties of a quantum system.

Disregarding the  origins of randomness in diverse settings, we shall take the
view   that stochastic processes are mathematically appropriate models when the time
evolution (parametric "dynamics" being  included) of random phenomena is involved.
Whenever probability laws are in usage, random phenomena and stochastic processes are
always at hand, \cite{lasota,sato}.

The major difference of our strategy, if compared to other approaches,
amounts to  considering exclusively  the parametric evolution (relaxation)
towards equilibiria of nearest neighbor spacing distributions as the
major source of probabilistic
information. We arrive here at prototype invariant measures of  limiting stationary
stochastic processes. We  do not invoke any explicit eigenvalue
(e.g. a solution of the spectral problem for the quantum system
or the related random-matrix model) nor level dynamics input, since those  data prove
to be  irrelevant for the primary ergodic behaviour that is displayed by the
adjacent spacing probability densities. At least in the considered approximatio regime,
where probability densities surmised by Wigner are assumed to be adequate (in reality,
they have the  status of reliable approximate formulas).

 An exploitation of  ergodicity (in fact strong mixing and/or exactness,
 \cite{lasota,mackey,rudnicki})
 of certain (parametrically evolving) Markovian stochastic  processes is here
found  to provide  a supplementary (probabilistic)
characterization of  quantum signatures of chaos.

\section{Poissonian level sequences}

\subsection{Exponential random variable and semi-Poisson laws}

"Poissonian"  matrix ensembles with independent random diagonal elements
are often used to model spectral properties of integrable Hamiltonian
systems (we disregard an issue of various, even quite remarkable,
 deviations from an  exact Poisson-type statistics, \cite{casati,marklof,veble}).
Indeed, many regular (integrable) systems,  semiclassically  exhibit
spacings between adjacent
energy levels which are distributed according to the exponential probability
density  $p(s) = exp(-s)$ on $R^+$, where we tacitly assume a normalization
of the first moment  (mean spacing) of the probability measure (hence the
unfolding of the energy spectrum, \cite{haake,veble}).

A canonical statement in this respect, \cite{berry}, conveys a  message that "for generic
regular systems" $p(s)$ is "characteristic of a Poisson process with levels distributed at
random"  and "the levels are not correlated". (A discussion of serious violations of the
Berry - Tabor
conjecture can be found in Ref. \cite{marklof}.)

 Since the regular spectrum is perfectly deterministic and for each set of quantum numbers
the corresponding energy level is obtained from an explicit formula (via
Einstein-Brillouin-Keller semiclassical argument, or directly by solving the spectral
problem for e.g.  rectangular billiard) it  is far form obvious  that levels
may come as a realization of a random variable. Even though  probability distributions
are thought to arise in near classical quantum systems, when the number of levels in
any range of energy is very large (and indefinitely increases when the  classical
limit is approached).

Following Refs. \cite{bohigas,veble}, let us  consider  a sequence of
numbers (we keep an explicit energy notation, although an  unfolded
sequence is rescaled to be non-dimensional):
\begin{equation}
E_{i+1} = E_i + x_i = E_0  + \sum_{j=1}^{i} x_j
\end{equation}
where $E_0=0$ and $x_j$ with $j=1,2,..$ are outcomes of independent
trials  of the exponentially distributed random variable $X$ taking values
in $R_+$.

The resulting  sequence $(E_1, E_2,...)$   of nonnegative numbers is a
particular model realization (sample) for what is commonly named a
Poisson spectrum. Here, randomly sampled (independent, in accordance with
the exponential distribution)
increments $x_i = E_{i+1} - E_i$ play the role of  adjacent level spacings.
Let us emphasize that the Poissonian random-matrix ensemble would comprise
 all  possible  sequences of the above form,  each obtained as a
result  of independent sampling procedures.

At this point let us turn to an  explicit probabilistic lore (cf.
\cite{feller,sato}) whose absence
is  painfully  conspicuous in major quantum chaos publications.

Let $X_1, X_2,...$ be independent random
variables with  common for all exponential probability law
$\mu (x)= \alpha exp(-\alpha x)$, $\alpha >0$ with mean $1\over \alpha $ and variance
${1\over {\alpha ^2}}$.

Furthermore let us denote $S_n = X_1 + X_2 +...+X_n$, $n=1,2,...$.
Then the   random variable  $S_n$  has a probability density:
\begin{equation}
p_{n}(x) = {{\alpha ^n\, x^{n-1}}\over {(n-1)!}}exp(-\alpha x)
\end{equation}
coming  from an  (n-1)-fold convolution of exponential probability
densities on $R^+$.
The law (2) is infinitely  divisibile, \cite{feller,sato}:
\begin{equation}
p_{n+m}(x) = (p_n*p_m)(x) = \int_0^x p_n(x-y) p_m(y) dy
\end{equation}
where   $p_1(x)=\mu (x)$  and $n,m = 1,2,...$.

In particular, note that  $X_i+ X_j$   for any $i,j, \in N$
has  a probability density
\begin{equation}
p_2(x)={\alpha ^2}\, x\,  exp(-\alpha x)\, .
\end{equation}
which upon setting $\alpha =2$ and $x=s$  stands for an example of
a semi-Poisson law  $P(s)=4s\, exp(-2s)$, \cite{haake,bogomolny},  which
has been identified to  to govern the  adjacent level
statistics for  a subclass of  pseudointegrable systems.

It is also
obvious that other (plasma-model related, \cite{bogomolny}) semi-Poisson laws
come  directly from distributions appropriate for  $S_n$.
For example,  $S_3$ has a density  $p_3(x)$ which upon substituting
$\alpha =3$ and $x=s$  gives rise to $P(s)={{27}\over 2}s^2 \, exp(-3s)$.
Analogously, $S_5$ yields  $p_5(x)$ and upon setting $\alpha =5$
implies $P(s)= {{3125}\over {24}} s^4\, exp(-5s)$, cf. Eq. (36) in Ref.
\cite{bogomolny}.

\subsection{Gaussian regime}

Both in the quantum chaos and random-matrix theory contexts, the
regime of $n>> 1$ is of utmost importance.
 Since  the primary random variable $X$ has an exponential density
with mean $\mu ={1\over \alpha }$ and variance  $\sigma  ^2 = {1\over \alpha ^2}$,
 we stay  within the   conditions of the central limit theorem, \cite{feller}.
First of all we know that for every $\epsilon >0$:
\begin{equation}
P[|{1\over n} S_n - \mu | > \epsilon ] \longrightarrow 0
\end{equation}
when $n\rightarrow \infty $.
Hence ${1\over n}S_n \rightarrow \mu $ with probability  $1$.\\
Furthermore,  we  have:
\begin{equation}
P[a< {{S_n - n\mu } \over {\sigma \sqrt{n}}} < b] \longrightarrow
{1\over {\sigma \sqrt{2\pi }}} \int_a^b exp[- {{(x-\mu )^2}\over {2\sigma ^2}}\,dx
\end{equation}

To give a pedestrian intuition about the above  formal observations,
let us ask for   a probability  that  there holds
\begin{equation}
|{S_n\over n} - \mu| < a {\sigma \over \sqrt{n}}
\end{equation}
for any $a>0$. In the regime of large $n$, an integral
${1\over {\sqrt{ 2\pi }}}\int_{-a}^a  exp(-y^2)\, dy$ gives a reliable answer.
The same integral determines the  probability that $|S_n - n\mu |< a\sigma \sqrt{n}$,
hence tells  us how $S_n$ fluctuates about $n\mu $  (and ${1\over n} S_n$ about $\mu $)
with the growth of $n$.

\subsection{Whence Poisson process on the energy axis ?}

The  probability density of the random variable $S_n$ allows  us to
evaluate a probability that the   $n$-th  level energy value $E_n$ is actually located
in an interval $[E, E + \triangle E]$ about a fixed nonnegative number $E$. It
is easily obtained by redefining the previous $p_n(x)$, cf. Eq. (2):
\begin{equation}
P[E \leq S_n \leq E+\triangle E] =
 {{\alpha ^n E^{n-1}}\over {(n-1)!}} exp(-\alpha E)\,  \triangle E =
\ee
 $$
 {{S^{n-1}}\over {(n-1)!}} exp(-S)\,  \triangle S = P_n(S) \, \triangle S
$$
where $x=E$, $S = {E\over { <E>}}$ and  ${1\over \alpha} = <E>$ is  the mean adjacent
level spacing.
The probability density   $P_n(S)$,
in Ref. \cite{haake}   is interpreted as  "probability density
for finding the $n$-th neighbor of a level in the distance increment
$[S,S+dS]$, for a stationary Poisson process", while in Ref. \cite{veble},  while
denoted $E(k,L)\rightarrow E(n-1,S)$  where $L$ is replaced by our $S$, stands for  the
"probability that inside an interval of length $S$ we find exactly $n-1$ levels".\\
Since $E(k,L)$  has the  form of a standard Poisson
probability law with mean-value  and variance $L$, one may also follow  \cite{veble}
to tell that "if they are on the average $L$ events, then the probability to actually
observe $k$ events is given by $E_{Poisson}(k,L) = {{L^k}\over {k!}}exp(-L)$".

Indeed, if $E>0$  is a  fixed
energy value  and we ask for a probability that there are exactly $n$ energy
levels below $E$, then  probability distributions  for $S_n$  and $S_{n+1}$
combine together to yield the Poisson distribution  with mean
$\alpha E$:
\begin{equation}
P[S_n \leq E < S_{n+1}] = P[N(E)=n] = {(\alpha E)^n \over {n!}} exp(-\alpha E)
\end{equation}

In this connection, let us recall that a random variable $N$ taking discrete integer values
$0,1,2,...$ is said to have Poisson distribution  with the mean  (and variance) $\lambda $
 if the probability of  $N=k$ reads $P[N=k]= {{\lambda ^k}\over {k!}} exp(-\lambda )$.
Clearly, $\sum_{k=0}^{\infty } P[N=k] = 1$ and
 $E[N] = \sum_0^{\infty } k P[N=k] = \lambda $.

Let us however stress that \it  no \rm explicit  Poisson process was  involved anywhere
in the  above,  since its
precise mathematical definition \cite{feller,billingsley} refers to a
counting process with a one parameter family of random variables
$ [N_t = N(t) = n] = [ S_n \leq t < S_{n+1} ] $  obeying  the
 Poisson probability law for all $t\in R^+$:
 \begin{equation}
P[S_n \leq t < S_{n+1}] =  P[N_t = n] = {{(\alpha t)^n} \over {n!}} exp(-\alpha  t) \, .
\end{equation}

The Poisson process has stationary independent increments:
 $N_{t_1},N_{t_2} - N_{t_1},...$ for $0<t_1<t_2<....$  with the Poisson
 probability distribution for each increment:
\begin{equation}
P[N_t - N_s = n] = {{[\alpha (t-s)]^n}\over {n!}} exp[-\alpha (t-s)]
\end{equation}
where $N_0=0$ with probability $1$.
Here, by denoting $P_n(t)=P[N_t=n]$ and $P_n(t-s)=P[N_t - N_s =n]$  we easily check  that
$\int_0^t P_n(t-s)P_m(s)(\alpha ds) = P_{n+m}(t)$.

 The  related intensity (parameter,  mean) of the Poisson "process"
equals $E[N_t] = \alpha t$  and displays the linear growth when $t$ increases.
Notice also that ${N_t\over t} \rightarrow \alpha $ with probability $1$ as
$t \rightarrow \infty $.\\
(The Poisson process is a particular example of a Markovian  process \it
in law \rm, \cite{sato}.
We deal here with  a temporally homogeneous process $N_t,\, t>0$ associated with an
infinitely divisible probability distribution $\mu (k) = {{c^k}\over {k!}}exp(-c)$,
The process in law is  here recovered  by simply setting
$\mu ^t(k) = {{(ct)^k}\over {k!}} exp(-ct)$ where $\mu ^1(k) = \mu (k)$.)

The Poisson process involves  time
dependent probabilities: $P_0(t) = exp(-\alpha t)$, $P_1(t) = \alpha t \, exp(-\alpha t)$,
..., which
 should be compared with  previous outcomes  for  the exponential
random variable.
By recalling Eq. (2)
we immediately arrive at a formal identification of probability distributions:
\begin{equation}
 p_{k+1}(t) = \alpha \, P_k(t)\, .
\end{equation}

In the above, the  exponential probability desnity  is labelled by time $t$.
Let us stress that
$p_{k+1}(t)\, \triangle t$ stands for a probability that the random variable
$S_{k+1}$  takes its  value in the  interval $[t, t+\triangle t]$, while
 $P_k(t)$  is a probability that $N_t=n$.

Notwithstanding,  Eq. (9) is formally identical with Eq. (10), and  therefore
we can in principle vary the parameter  $E$, so setting  (9) in a direct
equivalence with  a   parametric  (evolving in fictitious time)
Poisson process. This formal equivalence underlies a Poisson process lore of
 the quantum chaos literature.

Instead of paying attention to
the  exponential probability rule which  is  responsible for the
 randomness of the  collection of "time" instants  on
$R^+,$ one is tempted to tell that it is the Poisson process which dictates those
rules of the game. The standard way of thinking refers to the "observation of
the number of signals recorded up to an instant $t$ (actually, number of jumps of $N_t$
or the number of levels that are below $E$), \cite{feller}.

\subsection{Ergodicity}

Sample paths of the Poisson process $N_t$ are nondecreasing functions of $t$ with integer
values. If we attempt to draw  a  sample path,  we begin from the value $N_t =0$  which
is maintained up to the time instant $S_1=t_1$ when the jump occurs to $N_t= 1$.
This value stays  constant up to
the time $S_2=x_1+x_2$. Then,  a new jump to $N_t=3$ occurs, and that value survives until
$S_3=x_1+x_2+x_3$  is sampled.  The sample path construction for the Poisson process
strictly  parallels  a time series construction  in terms of  points on $R^+$ at which
 jumps of $N_t$ occur.
Intervals between consecutive time instants  form the sequence $(x_1,x_2,x_3...)$
of  adjacent level spacings.

On the other hand, it is
 Eq. (1) which provides us with a concrete  sample  sequence of levels
$(E_1, E_2,...)$, drawn in accordance with the exponential probability law for
adjacent level spacings $x_i, i\geq 1$.
Thus , the  set of all   realizations of the random  variable
$E = (S_1,S_2,...)$   comprises   a statistical ensemble of sample
  sequences $\omega : E(\omega ) = {(E_1,E_2,...)}$.
In fact, those sequences exemplify the   Poissonian ensemble  of spectra.\\
(If we  set $\alpha =1$, then a connection with the standard
Poissonian reasoning in the random-matrix approach to
quantum chaos is immediate. A catalogue of various statistical measures
for the Poissonian  spectra can be found in \cite{bohigas}.)

If we would construct a histogram of  adjacent level spacings for a   \it single \rm
sequence $(E_1, E_2,,,)$ which was compiled in accordance with the exponential
distribution, the familiar Poissonian shape would be revealed.

As well, the very same picture would emerge if we
would randomly collect
and make a statistical analysis of various finite strings   of
neighboring energy levels,
like in case of the  so called nuclear data ensemble composition (there e.g. one
makes a compilation of   1407 data  points from 30 sequences of levels experimentally
 found for 27 different nuclei), \cite{bohigas,mehta}.

All that is connected with a  primitive at this stage notion  of \it ergodicity \rm of the
exponential  \it process \rm.

Namely, let us consider a   one-parameter  family $(X_n,\, n=1,2,...)$
 of exponential random variables as   a stochastic process with "discrete time".
 Since $X_n$ are independent random variables with the \it same \rm for all $n$
 probability distribution,
 then for any real function $(f: x \in R^+ \rightarrow f(x) \in R)$  such that
 $<f> = E[f(X_1)]$
 exists,  we have
 \begin{equation}
 lim_{n \rightarrow \infty } {1\over n} \sum_{k=1}^n \, f(X_k) = <f> =
  \int_{R^+} f(x)\, \mu (dx)
 \end{equation}
for all sample sequences $X(\omega )= (x_1, x_2, ...)$.  In that case    the
random sequence $X_n, n\geq 1$ is   known to be \it ergodic with respect to \rm  $f$.
That is a standard link between the "time average" and "ensemble average", which is here
accomplished by means of the exponential probability measure $\mu $.
Indeed, as often happens in the the context of stationary stochastic processes,
ergodicity property allows us to replace  an average over the set of \it all \rm
 realizations of the  process at a chosen  time instant, by the  time average
 evaluated along \it one \rm sample trajectory.

If we  consider $f(X_n) = X_n$ for all $n\geq 1$, then  the ergodicity
notion refers to limiting properties of ${1\over n} S_n$. Accordingly,
 in view of the law of large numbers (Eq. (5)), Eq. (13) holds true.

Presently, there is no wonder in the fact that single eigenvalue series of a
suitable integrable quantum system (like e.g the rectangle billiard of Refs.
\cite{berry,veble,keating,casati})  may be utilized
 to generate a statistical  information in (approximate, \cite{marklof})
 affinity with the ensemble statistics.
Numerical research involving
 e.g. about $10^6$ - $10^{19}$ levels for  the eigenvalue
series $E_{m,n} = m^2 + \gamma n^2$,
 $\gamma = \pi /3$, cf. \cite{veble}, allows to generate various statistical data.
The nearest neighbor spacing histograms  show a very close resemblance
to the exponential distribution curve, in agreement with the conjecture of
Ref. \cite{berry}.
Effectively,  the eigenvalue  sequence of the rectangle billiard can be
interpreted as  (in fact mimics) a sample  path $E(\omega )=(E_1, E_2,...)$ with  adjacent
spacings $x_i$ distributed  according to  the exponential law.

A  standard (Poissonian) way of thinking in this context,  refers to  an
 "observation of the number
of signals  recorded up to an instant
$t$" (actually, jumps of $N_t$ or number of levels that are below  $E$), \cite{feller}.
However, the sample path  $E(\omega )$ encodes also a complete information about
a sample path of the  involved  exponential process $X=(X_1,X_2,...)$.

 Our ergodic argument
is valid with respect to any chosen  sample path
$X(\omega ) = (X_1(\omega )=x_1, X_2(\omega )=x_2,...)$  of $X$. An ensemble average
is provided by $\int_{R^+} x\mu (dx) = {1\over \alpha }$ and that value is to coincide with
 $lim_{n\rightarrow \infty } {1\over n}\sum_{k=1}^n x_n$ irrespective of the particular
 choice of a sample path $\omega $ of the exponential process $X_n, n\geq 1$.

 Ergodicity property normally  embodies  the weakest form of
complications present  in the evolution   of  physical systems,
including those modeled by stochastic processes.
There is  a well established  catalog of irregular behaviours that the dynamics of any
type may exhibit and there are stronger types of irregularity than those connected
with ergodicity. A corresponding hierarchy  of irregularities refers to the properties
of mixing and exactness \cite{mackey,rudnicki} which will be exploited
in below.

\section{Gaussian universality classes: Generalities}

In the random matrix approach  we have
a priori  involved   random-looking  sequences of energy levels, \cite{pastur},
which well agrees with the phenomenology of  nuclei  where inadequacies of
fundamental theoretical
models are  compensated by resorting to statistical matrix ensembles with
appropriate symmetries.  The roots of randomness presumably can be attributed to random
deformations of the "shape of the nuclei" (bag)  in the  independent-particle model
of nuclear dynamics,  \cite{jarzynski}. An  analogue of this reasoning can be
found in a recent analysis \cite{heller} of  a  chaotic  system in a
cavity (billiard) with a parametric   control of shape deformations. Then a
quantum particle is  confined within  a continuously deformed boundary , whose parametric
dynamics can be as well represented by a stochastic process of any kind.

A concrete quantum system (like e.g.  a spectral problem for concretely  shaped billiard)
usually induces its own unique spectrum and there
is no need, nor room  for any statistical ensemble  of systems (unless we shall
indeed consider
a  family of quantum systems with a suitable selection of random potentials.
  We must thus cope with  obvious discrepancies underlying    otherwise attractive
  affinities (e.g. the universality  classes idea for spectral statistics).
Useful affinities appear to mask quite deep differences between the underlying physical
 mechanisms.

It is the level repulsion which is routinely  interpreted  as a quantum
manifestation of classical nonintegrability and ultimately also  of chaos,
cf. \cite{pechukas}.

 Normally that is quantified by means of polynomial  modifications  of the
 Gaussian probability law (in association with the Wigner-Dyson  statistics  of adjacent
 level spacings for  e.g. unitary, orthogonal and   symplectic  random matrix ensembles).
For completness of the argument, let us list the standard formulas:
$P_1(s) =    s {{\pi }\over 2} exp(-{{s^2\pi }\over 4})$,
$P_2(s) = s^2 {{32}\over \pi ^2} exp(-{{s^2 \pi }\over 4})$ and  $P_4(s) =
s^4 {{2^{18}}\over {3^6 \pi ^3}}\, exp(- {{s^2 64}\over {9\pi }})$,
corresponding respectively to the GOE, GUE
and GSE random-matrix theory predictions.

Let us point out \cite{haake} that for most practical cases the Wigner distributions (albeit
exact in the $2\times 2$ random-matrix case only) are adequate. Typical spacing histograms
drawn from experimental or numerically generated (quasi)energy spectra are too rugged to
allow  subtle distictions against the $n\rightarrow \infty $ random-matrix size related
predictions.

We shall consider  mostly the   Wigner-type  cases, even though  neither
  of those probability laws deserves the status of being an \it exact \rm
  representation of the real state of affairs.
Remember that also in the context of random matrix theories the Wigner spacing formulas  are
 approximations that usually improve in the large matrix size regime.

The nearest  neighbor spacing distributions, in the random-matrix  approach are the
secondary notions and can be derived from an explicit formula for the   joint
probability density  to find the (dimensionless) energy eigenvalues
in respective  infinitesimal  intervals $[x_i, x_i + \triangle x_i]$ with $i=1,2,...,N$:
\begin{equation}
P(x_1,x_2,...,x_N) = C_{N\beta } \, [\prod _{i>j=1}^N |x_i - x_j|^{\beta } \,
exp(- {1\over 2}\sum_{i=1}^N x_i^2)]
\ee
where $\beta =1,2,4$ and $C_{N\beta }$ is a normalization constant, \cite{haake,mehta}.
The  level repulsion has been built into the framework from the
very beginning and appropriate level spacing distributions (including the adjacent
level case) can be directly evaluated on that basis, \cite{mehta,shukla,guhr}.

There were many attempts to provide convincing (and independent
from the definite symmetry and Gaussian randomness inputs,
proper to random-matrix theory)) arguments that  would
generate level  repulsion   through  well defined dynamical mechanisms
(like e.g. the  parametric level dynamics) and would lead to statistical
predictions as well.  A suitable level dynamics scenario may as well
give rise to  the so-called intermediate statistics and possibly a continuous
(parametric) interpolation among them.

 In the random matrix theory context a radical  probabilistic   attempt due to
 Dyson explicitly
 involves the (parametric) Brownian motion assumption for each energy level
 separately, \cite{mehta,guhr,dyson}.

More satisfactory results were obtained by resorting to a fictituous gas of
interacting particle representatives of individual energy levels. A corresponding
many-particle  system  is then investigated at suitable "thermal equilibrium" conditions.
Then, without introducing a priori statistical ensembles of random matrices,
level distribution functions are derived by means of ordinary statistical mechanics methods.
That approach explictly involves the many-body Hamiltonian (Calogero model):
\be
H= -  \sum_{i=1}^n {{\partial }^2\over  {\partial x_i}^2} + {\beta (\beta -2)\over 4}
\sum_{i<j} {1\over {(x_i-x_j)^2}} + \sum_{i=1}^n x^2_i
\ee

whose squared ground state  function (equlibrium measure density) has the form (14),
\cite{guhr,shukla}.

Apart from that, explicit quantum mechanical investigations for billiard-type systems
provide hints about the potential importance of  interpolation studies, especially since
various intermediate types of statistics were reported to occur, see e.g.
\cite{stock,heller,hu}.

There are two basic approaches to an interpolation issue.
One refers explicitly to random matrix theories and their "affinity" with
quantum chaotic systems, \cite{hasegawa,aberg,shukla}. Another refers to the
fictitious gas, interacting many-body analogy,
\cite{pechukas,yukawa,lenz,hasegawa1,stock1,haake}.
Recently, a related short-range plasma model was proposed to analyze an emergence of the
"pseudo-Poisson statistics", \cite{bogomolny}.

\section{Parametric dynamics of adjacent level spacings}

\subsection{Markov processes defined through their invariant measures}

Once  we have encountered probability densities on the positive
half-line in $R^1$, it is rather natural to investigate a general
issue  of   parametric stochastic processes which would provide  a
dynamical model  of   level repulsion in an irregular quantum
system and generate at the same time   spacing densities as those
of  asymptotic invariant (equilibrium)  probability measures. Such
random processes clearly must  run with respect to the previously
mentioned "fictitious" time-parameter and take values in the set
of all  level spacings  which are appropriate for  a complex
quantum system or the corresponding random-matrix ensemble.

Effectively, we wish to introduce a  Markovian diffusion-type
process which might stand  for a reliable approximation of  a
random walk over \it  level spacing sizes.\rm

For future reference let us mention that in the regime of
equilibrium (when an invariant measure appears
 in the large "time"  asymptotic), a sample path of such random walk would take the form of
 an ordered sequence of spacings which are sampled (drawn) according to the prescribed
  invariant probability distribution.
 That is precisely  \it one  \rm explicit example of the  ladder  of energy levels,
 understood as a random sample drawn from a suitable ensemble.

 An analysis of statistical features
of this  spectral sequence  involves an ergodicity notion to stay
in   conformity with the ensemble  evaluation of various averages
(carried out with respect to the invariant density),
\cite{lasota,lefever}.

We shall consider the previously listed  GOE, GUE and GSE
probability densities on $R^+$  (up to  suitable rescalings !) as,
 distorted in view of the spacing size normalization,
asymptotic invariant densities  of certain parametric Markovian
stochastic processes whose uniqueness status can be unambiguously
settled.

Let is begin from the observation that probability densities  on
$R^+$, of the characteristic form $f(x) \sim x exp(-{x^2\over
4})$, $g(x) \sim  x^2 exp(-{x^2\over 2})$ and
 $h(x)\sim {x^4\over 4} exp(-x^2)$
appear notoriously in various quantum mechanical contexts
(harmonic oscillator or centrifugal-harmonic  eigenvalue
problems), cf. \cite{calogero,zambrini,blanch,cufaro,cufaro1}.
Notwithstanding, as notoriously they can be identified in
connection with special classes of stationary Markovian diffusion
processes on $R^+$, \cite{karlin}.

Anticipating further discussion, let us consider a Fokker-Planck
equation on the positive half-line in  the form: \be
\partial _t\rho  = {1\over 2} \triangle  \rho   - \nabla [{{\beta }
\over {2x}} - x)\rho ] \ee which may be set in correspondence with
the stochastic differential equation $ dX_t = ({{\beta }\over
{2X_t}} - X_t)dt + dW_t \, $ formally valid for a random
variable   $X_t$  with values contained in $(0,\infty )$. Here
$\beta \geq 0$ and $W_t$ represents the Wiener process.

Accordingly,  if $\rho _0(x)$ with $x\in R^+$  is regarded as  the
density of distribution of $X_0$ then for each $t>0$ the function
$\rho (x,t)$, solving Eq. (16), is the density of $X_t$. In view of
a singularity of the forward drift at the origin, we refrain from
looking for strong solutions of the  above stochastic differential
equation and confine attention to weak solutions only and the
associated tractable parabolic problem (16q) with suitable boundary
data, cf. \cite{karlin}.

In all those cases a mechanism of repulsion is modeled by the
$1\over x$  term in the forward drift expression. The compensating
harmonic attraction  which is    modeled by the $-x$ term,
saturates the long distance effects of repulsion-induced
scattering   and ultimately yields asymptotic steady (stationary)
probability densities.

To interpret a  density   $\rho (x)$ as an asymptotic (invariant)
density of a well defined  Markovian diffusion process  we shall
utilize the rudiments of  so-called Schr\"{o}dinger boundary and
stochastic interpolation problem,
 \cite{zambrini,blanch,garb}, see also \cite{pinsky} when
 specialised to invariant measures.

 Let us notice that both in case of the standard Ornstein-Uhlenbeck process
 and its Bessel (radial)  variant, we have emphasized the role of
 a stochastic process with an asymptotic invariant density.
 To deduce such processes,  in principle we can   start from
  an invariant density and  address an easier issue of the
    associated measure preserving stochastic dynamics. Next we can consider
      whether the obtained    process  would drive  a  given initial density towards
   a prescribed invariant measure (in that case we can tell about an asymptotic state of
    equilibrium to which the process relaxes).
    That feature involves the notion of  exactness of the
   related stochastic process, whose straightforward consequence are the properties of
    mixing and ergodicity of the corresponding   random dynamics, \cite{lasota}.

  There is a general formula \cite{pinsky,zambrini,garb} relating the forward drift of
  the  sought for stationary process
  with an explicit functional form of an invariant probability density.
  We confine our attention to
  Markov diffusion processes with a constant diffusion coefficient,  denoted
  $D >0$. Then, the pertinent  formula reads:
  \be
  b(x) =  2D {{\nabla \rho ^{1/2}}\over {\rho ^{1/2}}}\, .
  \ee

 In particular,  for the familiar Ornstein-Uhlenbeck process
 we have  $\rho ^{1/2} (x) =
({1\over {\pi  }})^{1/4} exp(-{x^2\over 2})$ and  $D= {1\over 2}$,
so
 we  clearly arrive at  $b(x)= - x $ as should be.
Quite analogously,  in case of  the GUE-type  spacing density, we
have $D ={1\over 2}$ and $\rho ^{1/2}(x) = {2\over {\pi ^{1/4}}} x
\exp(-{x^2\over 2})$. Thus, accordingly $b(x)= {1\over x} - x$.

The very same strategy allows us to  identify a forward drift of
the Markovian diffusion process supported by the GOE-type spacing
density. By  employing   $\rho ^{1/2}(x)= \sqrt {2x} exp(-
{x^2\over 2})$ and setting $D = {1\over 2}$ we  arrive at the
formula: $b(x,t)= {1 \over {2x}} - x $.

We immediately identify the above forward drifts with the ones
appropriate for  the time homogeneous  radial  Ornstein-Uhlenbeck
processes, with a corresponding family of ($N>1$ and  otherwise
arbitrary integer) transition probability densities,
\cite{karlin}: \be p_t(y,x) = p(y,0,x,t) = 2 x^{N-1}
\exp(-x^2)\cdot \ee
$$
{1\over {1-\exp(-2t)}}  \exp[-{{(x^2 + y^2)\exp(-2t)}\over
{1-\exp(-2t)}}]\cdot
$$
$$
[xy\exp(-t)]^{-\alpha } I_{\alpha }({{2xy\exp(-t)}\over
{1-\exp(-2t)}})
$$
where $\alpha = {{N-2}\over 2}$ and $I_{\alpha }(z)$ is a modified
Bessel function of order $\alpha $: \be I_{\alpha } (z) =
\sum_{k=0}^{\infty } {{(z/2)^{2k+\alpha }} \over { (k!) \Gamma
(k+\alpha + 1)}}
 \ee
while the Euler gamma function has a standard form $\Gamma (x) =
\int_0^{\infty } \exp(-t) t^{x-1} dt$.   We remember that $\Gamma
(n+1) = n!$ and $\Gamma (1/2) = \sqrt {\pi }$.

 The  resultant forward drift has the general form:
\be b(x) = {{N - 1}\over {2x}} - x\, \ee and corresponds to $\beta
= N-1$.

By setting $N=2$, and then  employing the series
representation of $I_0(z)$, we easily  recover the asymptotic
invariant density for the process: $lim_{t\rightarrow \infty }
p(y,0,x,t) = 2x\exp(-x^2)$.

We can also analyze the large time asymptotic of $p(y,0,x,t)$,
 in case of $N=3$
 which gives rise to  an  invariant  density in the form:
${4\over {\sqrt{\pi }}} x^2 exp(-x^2)$. That obviously corresponds
to the GUE-type case with $b(x)= {1\over x} - x$.

When passing to the GSE case, we are interested in the Markovian
diffusion process which is supported by an invariant probability
density $\rho (x)= {2\over {\Gamma (3/2)}} x^4 \exp(-{x^2})$. Let
us evaluate the forward drift of the sought for process
(we set $D = {1\over 2}$): $b(x,t) =
{2\over x} - x$.
Clearly,  we  deal  here with a
radial  Ornstein-Uhlenbeck process corresponding to $N=5$.
The transition probability density of the process
displays an expected asymptotic: $lim_{t\rightarrow \infty }
p(y,0,x,t) = {4\over {\sqrt{\pi }}} x^4 \exp(-x^2)$. Here we have
exploited $\Gamma (1/2) = \sqrt{\pi }$ to evaluate $\Gamma (3/2)=
{1\over 2}\sqrt{\pi }$.

The above formulas   allow us  to formulate  a hypothesis
that further  non-generic repulsion laws  may be  appropriate for
quantifying quantum chaos. Straightforwardly,  one can verify that
our transition probability densities  refer to asymptotic
invariant densities of the  form:
 \be \rho (x) =  {2\over {\Gamma (N/2)}} x^{N-1} exp(-x^2)
 \, . \ee

   In particular we get a direct  evidence in favor of   $N = 4$,
i. e. $b(x) = {3\over {2x}} - x $,  universality class which in
fact corresponds to the Ginibre ensemble of of  non-Hermitian
random matrices, \cite{haake}, where
 a cubic level repulsion appears: $\rho (x)=2 x^3 \exp(- x^2)$
 (this formula is exact for  $2\times 2$ random matrices).

In principle,   processes
corresponding to any  $N > 5$ may  be realizable as well, and thus
the related higher-power level repulsion  might have relevance  in
the realm of quantum chaos.

In all considered cases, an asymptotic invariance of probability
measures (densities)
 is sufficient to yield  ergodic behaviour.
 For each value of $N>1$ we deal
with an independent repulsion mechanism, albeit all of them belong
to the radial Ornstein-Uhlenbeck family.

We have thus  identified a universal stochastic law (in fact, a
family of the like) behind the functional form of basic,
Wigner surmise inspired, spacing
probability densities appropriate for  quantum chaos.

Let us emphasize at this point that one should keep in mind a
number of  possible
 reservations coming from the fact that neither of "universal"  or "generic"
 laws  can be regarded as a faithful representation  of a real state of affairs.
 Usually exact laws are derived for two by two (hence of the small size !)
 random  matrices, and are known to  reappear again (at least in the generic cases)
 as approximate spacing formulas in the large random-matrix size regime. That in
 turn allows to achieve a correspondence with semiclassical quantum spectra
 of complex systems.

There is no obvious  explanation of a physical meaning of the
integer parameter $N$ in the radial stochastic process scenario.
One hypothesis comes from the random-matrix theory, where $\beta =
N-1 = 1, 2, 4$ would correspond to  a number of  independent
components of a typical matrix entry which is decided by the underlying symmetry
of the problem (GOE, GUE, GSE).
That can be presumably be  extended to the case
of $N=4$  and possibly all $N\leq 8$.

\subsection{Link with Calogero Hamiltonian}

Previously we have indicated  that a common mathematical basis for
various level repulsion mechanisms appropriate to quantum chaos is
set by the Calogero-Moser Hamiltonian.
  At the first glance, our stochastic arguments may leave an impression that
something completely divorced from that setting has been obtained
in the  present paper. However things look otherwise  and our
theoretical framework proves to be compatible with standard
techniques for spectral analysis of complex quantum systems.

It is peculiar to the general arguments  of Refs.
\cite{zambrini,garb} that  invariant probability densities give
rise to measure preserving stochastic processes in a fully
controlled way. One of  basic ingredients of the  formalism is a
proper choice  of  Feynman-Kac kernel functions, which are the
building block for the construction of
 transition probability densities of the pertinent Markov processes.
  Feynman-Kac semigroup operators (and their kernels) explicitly  involve
one particle Hamiltonian operators as generators (in less
technical terms one may think at this point about  rather
standard transformation from the Fokker-Planck operator to the
associated self-adjoint one, \cite{risken}).

For stationary processes, a  general formula relating forward
drifts $b(x)$ of the stochastic process with potentials of the
conservative Hamiltonian system reads (we choose a diffusion
coefficient to be equal $1\over 2$), \cite{blanch,garb}: \be V(x)
= {1\over 2} (b^2 + \nabla \cdot b)\, . \ee

Upon substituting the general expression for  $b(x)$
 we arrive at:
 \be
V(x) = {1\over 2} [ {{\beta (\beta -2)}\over {4 x^2}} - (\beta +1)
+ x^2] \ee where $\beta = N-1$. This potential function enters a
standard definition of the one particle Hamiltonian operator
(physical parameters have been scaled away): \be H = -{1\over 2}
\triangle   + V(x) \ee where $\triangle = {{d^2}\over {dx^2}}$.
The operator (24) with $V(x)$ defined by (23) is an equivalent form
of  a two-particle  (actually two-level) version of the
Calogero-Moser Hamiltonian, cf. \cite{calogero}.

Indeed, the classic Calogero-type problem is defined by \be H = -
{1\over 2}{d^2\over {dx^2}} + {1\over 2}x^2 + {{\beta (\beta - 2)}
\over {8x^2}} \ee
with the well known spectral solution. The
eigenvalues read $E_n(\beta ) = 2n + 1 + {1\over 2}[1+\beta (\beta
- 2)]^{1/2}$, where  $n\geq 0$ and $\beta > -1$.

By inspection we can check that all previously considered $N=
2,3,4,5$ radial processes correspond to the  Calogero  operator of
the form  $H - E_0$ where $E_0$ is the ground state  (n=0)
eigenvalue. Its explicit foom relies on the choice of $\beta $ and
by substituting $\beta = 1,2,3,4$ we  easily check that
\be
E_0(\beta
) = 1+ {1\over 2}[1+\beta (\beta - 2)]^{1/2}= {1\over 2}(\beta +
1)\, .
\ee

Accordingly, all  considered radial processes arise as  the
so-called ground state processes associated with the Calogero
Hamiltonians  (squared modulus of the ground state wave function
stands for the pertinent probability density). Let us recall  that
the classic Ornstein - Uhlenbeck process can be regarded as  the
ground state process of the harmonic oscillator Hamiltonian
operator. That by the way corresponds to choosing $N=1$ i.e.
$\beta  = 0$ in the above, plus allowing the whole of $R^1$ to the
process. Like in the standard OU process case, radial OU processes
share the property of exactness (while driving any initial density
towards suitable equilibrium) and hence ergodicity.

\section{Discussion}

Our motivations were  essentially probabilistic and spectral series with spacing densities
governed by Wigner-type laws have emerged in the course of a parametric stochastic
process that relaxes towards equilibrium (invariant measure). Such series have thus a
definitely random origin. It is clear that an approximate value of Wigner densities
indicates nonrandom input in realistic cases.

Let us point out that in the standard matrix-theory framework  Dyson' s "threefold way"
is based on the demonstration that on general (invariance under symmetry) grounds only
three basic  ensembles
(orthogonal, unitary and symplectic) matter. Hence, the non-generic repulsion
behaviour we have discussed before, goes beyond the standard framework (under assumptions of
the Dyson theorem the non-generic laws  are not admissible).
Many different ensembles  have been used in the
literature, but their properties were more specific (less general) than the standard GOE,
GUE and GSE cases show up.

The spacing distributions we have addressed (Wigner surmise),  fail to be correct in general.
 The true random-matrix universal distributions differ from them, albeit the discrepancy
is small when matrix size is going to infinity, \cite{haake}.

In the discussed parametric relaxation process scenario, one may easily implement a transition
of any initial density towards a concrete asymptotic one with the wealth of intermediate
examples (e.g. from Poisson to GOE interpolation).
In that case, both the initial and terminal distributions refer to random seqences of
 numbers (possible energy eigenvalues).
Ergodicity of the ultimate stationary process implies that its sample paths arise as
random sequences drawn from the Wigner-type distribution. Clearly, we cannot expect that such
purely random sample sequences would reveal long range correlations typical of
random-matrix models.

{\bf Acknowledgement:} I would like to thank Karol \.{Z}yczkowski for correspondence  on
semi-Poisson laws and for pointing Ref. \cite{marklof} to my attention .

\end{document}